\documentstyle[seceq,twoside,epsf,wrapft]{ptptex}
\notypesetlogo  
\title{Observed Property of $\sigma$-Meson and Chiral Symmetry}
\author{%
Muneyuki {\sc Ishida} }
\inst{%
Department of Physics, Tokyo Institute of Technology\\
Tokyo 152-8551, Japan
}
\abst{%
The phenomenologically observed property of $\sigma$(600) in our recent 
analyses of various $\pi\pi$-scattering and -production
processes is reviewed and compared with the prediction of $SU(2)$ linear 
$\sigma$ model. Furthermore, a possibility of the scalar $\sigma$-nonet 
($\sigma (600)$, $\kappa (900)$, $f_0(980)$ and $a_0(980)$) forming 
with the pseudoscalar $\pi$-nonet a linear representation of $SU(3)$ chiral
symmetry is investigated. The origin of repulsive background phase shift 
$\delta_{BG}$, which is essential to lead us to the $\sigma$-existence 
in our phase shift analysis, is shown to come from the repulsive $\lambda\phi^4$
interaction.
}

\begin{document}
\maketitle

\setcounter{tocdepth}{4}

\section{Introduction}
Recently we have 
found rather strong evidence for existence of the light 
$\sigma$-meson through a series of analyses of $\pi\pi$-scattering 
and various $\pi\pi$-production processes. 
First we collect the values of obtained mass and width of $\sigma$-meson.
We also review our consideration, whether the $\sigma$-meson with this property
deserves to be the member of a linear representation of chiral $SU(3)$ symmetry or not.
Secondly we explain that the origin of repulsive background phase shift 
$\delta_{BG}$, introduced in our phase shift analysis and led to 
$\sigma$-existence,
is compensating $\lambda\phi^4$ interaction, which is necessary from the viewpoint of 
chiral symmetry. 
Finally 
comparison of our method of phase shift analysis
with the other methods, which did or did not lead to $\sigma$-existence, 
is given in some detail.

\section{Observed Property of $\sigma$-Meson and Linear $\sigma$ Model}
\hspace*{-0.7cm}\underline{\em Observed\ Properties\ of\ $\sigma$-Meson}\ 
obtained through in our recent analyses of the various processes
are given in Table I. 
\begin{table}
\begin{center}
\caption{Observed mass and width of $\sigma$-meson in our analyses}
\begin{tabular}{|l|c|c|}
\hline\hline
Process & $m_\sigma$/MeV & $\Gamma_\sigma$/MeV \\
\hline
\underline{Present\ Our\ Estimate} & \underline{500$\sim$675} & 
\underline{300$\sim$500}\\
$\pi\pi$-scattering\cite{rf1} & 535$\sim$675 & 385$\pm$70 \\
($pp$ central collision\cite{rf2} & 590$\pm$10 & 710$\pm$30) \\
$J/\psi\rightarrow\omega\pi\pi$ decay\cite{rf2} & 480$\pm$5 & 325$\pm$10 \\
$p\bar p \to 3\pi^0$\cite{rf4} & 540$\stackrel{+36}{\scriptstyle -29}$ &  
385$\stackrel{+64}{\scriptstyle -80}$ \\
$\Upsilon^{(m)}\to\Upsilon^{(n)}\pi\pi$,$\psi\to J/\psi \pi\pi$\cite{rf5}
 & 531$\pm$  & 281$\pm$ \\
\hline
\end{tabular}
\end{center}
\end{table}

\hspace*{-0.7cm}\underline{\em SU(2)L$\sigma$M}\ \ \ \ 
In the SU(2) linear $\sigma$ model   
the coupling constant $g_{\sigma\pi\pi}$ of the 
$\sigma\pi\pi$ interaction is related to the $\lambda$ of the $\lambda\phi^4$ 
interaction
and $m_\sigma$ as
\begin{eqnarray}
g_{\sigma\pi\pi} &=& f_\pi\lambda =(m_\sigma^2-m_\pi^2)/(2f_\pi ).
\label{eqrel}
\end{eqnarray}
Thus, the width of $\sigma$-meson $\Gamma_\sigma$ is related with its mass
$m_\sigma$ through the relation   
$\Gamma_{\sigma\pi\pi}^{\rm theor} = 
\frac{3g_{\sigma\pi\pi}^2}{4\pi m_\sigma^2}p_1, 
\ \ p_1=\sqrt{\frac{m_\sigma^2}{4}-m_\pi^2}$ .
This relation between $m_\sigma$ and $\Gamma_\sigma$ is not derived 
from the non-linear treatment of $\sigma$ meson. 
By using this relation with our presently estimated value of $m_\sigma$ 
in Table I
we can predict the value of $\Gamma_\sigma$ as 
$\Gamma_\sigma^{\rm theor}=300$MeV(for $m_\sigma$=500MeV)$\sim$900MeV
(for $m_\sigma$=675MeV), which is consistent\cite{rf6} with our present experimental 
estimate of $\Gamma_\sigma$ given in Table I. 
Thus, the observed $\sigma$ meson may be identified with the $\sigma$
meson described in the L$\sigma$M. \\
%
\underline{\em SU(3)L$\sigma$M}\ \ \ \  
Through the analysis\cite{rf7} of $K\pi$-scattering phase shift by a 
similar method to $\pi\pi$, we have obtained an evidence of 
existence of the $I=1/2$ 
scalar $\kappa (900)$-meson.  

Now we have the scalar mesons below 1 GeV, $\sigma (600)$, $\kappa (900)$, 
$a_0(980)$ and $f_0(980)$. Previously I discussed that the properties of these 
scalar mesons are consistent\cite{rf8} with those  
predicted by SU(3)L$\sigma$M, as shown in 
Table II.\footnote{The predicted properties are very sensitive to the ratio of 
$f_K/f_\pi$. We prefer the region of this ratio  
$1.329<f_K/f_\pi<1.432$ (somewhat larger than the experimental 
value 1.22), where the value $m_\kappa^{\rm theor}$ reproduces the 
experimental value within its uncertainty.}

\begin{table}
\begin{center}
\caption{The properties of the scalar meson nonet
predicted by $SU(3)$L$\sigma$M, compared with experiments. 
The underlined values of $m_\sigma$ and $m_\kappa$
along with $f_\pi$, $m_\pi$, 
$m_\eta$ and $m_{\eta '}$ are used as inputs, and 
we can predict the masses and widths of scalar mesons.
 }
\begin{tabular}{|l|c|c|c|c|}
\hline
\hline
   & $m^{\rm theor}$/MeV & $m^{\rm exp}$/MeV
   & $\Gamma^{\rm theor}$/MeV & $\Gamma^{\rm exp}$/MeV\\
\hline
  $\sigma$ & $\underline{535\sim 650}$ & $\underline{535\sim 650}$
           & $400\sim 800$       & $385\pm 70$ \\
\hline
  $\kappa$ & $\underline{905\stackrel{+65}{\scriptstyle -30}} $
           & $\underline{905\stackrel{+65}{\scriptstyle -30}}$  
   & $300\sim 600$ & $545\stackrel{+235}{\scriptstyle -110} $\\
\hline
  $\delta =a_0(980)$ & $900\sim 930$ & $982.7\pm 2.0$  
           & $110\sim 170$ & 
$95\pm 14$ \\
\hline
  $\sigma '=f_0(980)$ & $1030\sim 1200$  & $993.2\pm 9.5$
             &  $0\sim 300$   & $67.9\pm 9.4$\\
\hline
\end{tabular}
\label{tab:tree}
\end{center}
\end{table}

Being based on these results,
it may be plausible to regard ${\sigma}${\bf (600)}, 
${\kappa}${\bf (900)}, {\bf a}${}_0${\bf (980)}, {\bf and} 
{\bf f}${}_0${\bf (980)}{\bf as members of the scalar nonet,
forming with the members of} ${\pi}${\bf -nonet a linear 
representation of the SU(3) chiral symmetry}.\footnote{
This interesting assignment was suggested and insisted upon repeatedly by 
M. D. Scadron.\cite{rf9} 
}
Here it is also to be noted that the octet members of this scalar nonet,
$\kappa ,\  a_0$ and $\sigma_8$(mixture of $\sigma$ and $f_0$), 
satisfies\cite{rf8} the Gell-Mann Okubo mass formula.

\section{Origin of Repulsive Phase and Comparison with Other Analyses}
\hspace*{-0.7cm}\underline{\em Origin of the $\delta_{BG}$ }\ \ \ 
An important physical reason, which led us to $\sigma$-existence, 
is introduction of the background phase 
$\delta_{BG}$\footnote{This 
strong repulsive phase shift is cancelled out by the attractive
contribution from $\delta_\sigma$ in $I=0$ channel, and not observed directly. 
However, in $I=2$ channel there is no known/expected resonance, 
and the repulsive phase shift is directly observed experimentally,
and can be fitted\cite{rf1} well by the hard core formula $\delta^2 =-p_1r_c$ with core radius
$r_c$=0.87GeV$^{-1}$ (0.17fm). 
} 
in our phase shift analysis.
The origin of this $\delta_{\rm BG}$ is considered to have a close connection to 
the $\lambda\phi^4$-interaction in L$\sigma$M\cite{rf6} as follows:
The $I=0$ $\pi\pi$-scattering amplitude is given by 
$3A(s,t,u)+A(t,s,u)+A(u,t,s)$.  The $s$-channel term, 
$3A(s,t,u)$, is main contribution, which is given by $SU(2)$L$\sigma$M as 
\begin{eqnarray}
3A(s,t,u) &=& 3(-2g_{\sigma\pi\pi})^2/(m_\sigma^2-s)-6\lambda 
\equiv {\cal K}_\sigma +{\cal K}_\lambda,
\label{eqamp}
\end{eqnarray}
where we define, for later convenience, 
the 1st term as ${\cal K}_\sigma$ and the 2nd term as ${\cal K}_\lambda$.
Because of the relation (\ref{eqrel}), the dominant part 
due to $\sigma$-resonance (1st term) is cancelled by that due to repulsive 
$\lambda\phi^4$ interaction (2nd term) in $O(p^0)$ level:
\begin{eqnarray}
3A(s,t,u) = \frac{3}{f_\pi^2}\frac{(m_\sigma^2-m_\pi^2)^2}{m_\sigma^2-s}
-\frac{3(m_\sigma^2-m_\pi^2)}{f_\pi^2}
=3\frac{s-m_\pi^2}{f_\pi^2}+\frac{3}{f_\pi^2}
\frac{(m_\pi^2-s)^2}{m_\sigma^2-s},\ \ \ \ \ \ \ 
\label{eq:Acancel}
\end{eqnarray}
where in the last side the $O(p^2)$ Tomozawa-Weinberg amplitude
and the higher order correction term are left.
As a result the derivative coupling property of $\pi$ as
Nambu-Goldstone boson is preserved. 
This strong cancellation in L$\sigma$M correponds to that between
$\delta_\sigma$ and $\delta_{\rm BG}$ in our phase shift analysis. 
Thus the origin of $\delta_{BG}$ has been proved to be 
the ``compensating repulsive 
interaction" 
necessarily required from the viewpoint of 
chiral symmetry,\footnote{It should be noted that in the analysis\cite{rfkamin} by 
Kaminski et al., leading to the light $\sigma$-existence, this repulsive 
phase shift is implicitly introduced through the use of the Yamaguchi-type 
potential.}
as schematically shown in Fig. 1.

\begin{figure}[t]
\parbox{\halftext}{
  \epsfysize=5.5 cm
  \centerline{\epsffile{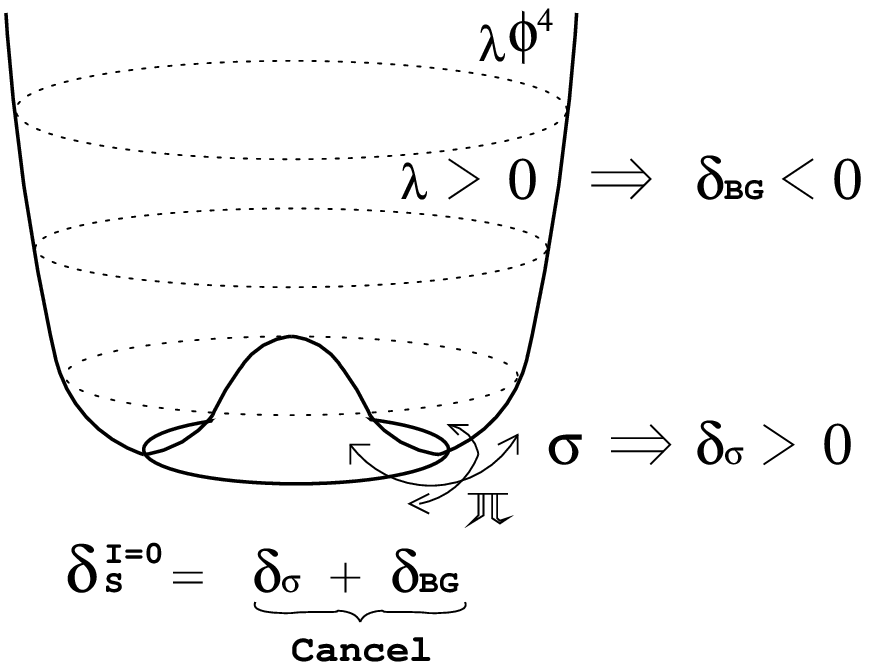}}
  \caption{``Mexican hat" potential, which may be considered 
as  QCD effective potential. 
The repulsive interaction between pions 
due to $\lambda\phi^4$ term in L$\sigma$M 
is the origin of negative phase $\delta_{BG}$.   
Total phase shift $\delta$ is given by $\delta =\delta_\sigma +\delta_{BG}$
($\delta_\sigma$ being due to $\sigma$ formation).
}
  \label{fig2}}
 \hspace{4mm}
 \parbox{\halftext}{
  \epsfxsize=6. cm
  \epsfysize=6. cm
  \centerline{\epsffile{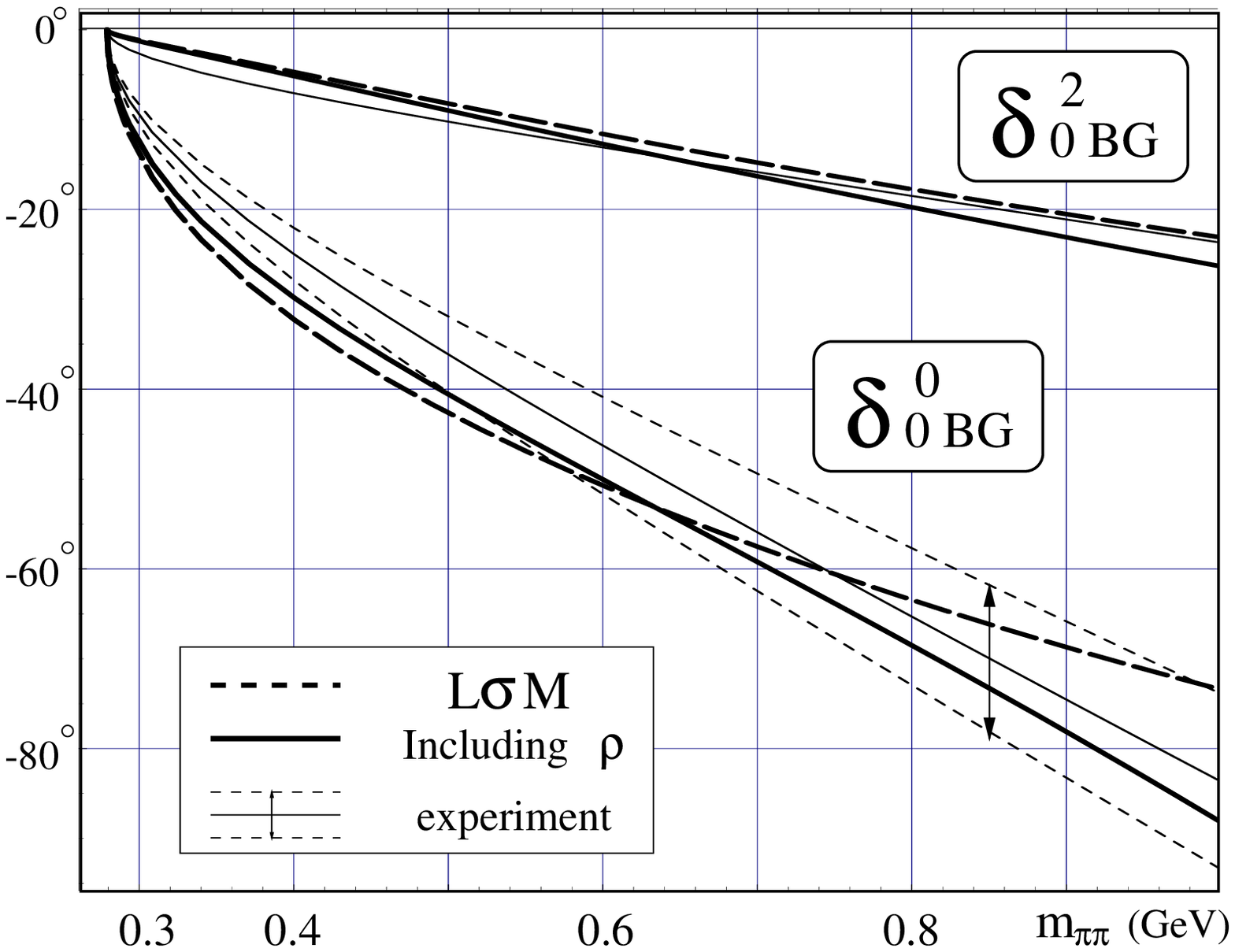}}
  \caption{The repulsive $\delta_{BG}$ in $I=0$ and $I=2$ $\pi\pi$ scattering 
           predicted by the L$\sigma$M with and without 
           $\rho$-meson contribution. The results are compared with the 
           phenomenological $\delta_{BG}$ of hard core type.}
  \label{fig3}\vspace{0mm}}
\end{figure}

In Fig.~2 I have shown our theoretical $\delta_{BG}$ estimated 
semi-quantitatively and compared with the phenomenological $\delta_{BG}$
of hard core type. We have unitarized  
the Born amplitude ${\cal K}_\sigma$ and 
 ${\cal K}_{BG}(={\cal K}_\lambda$ 
in Eq.(\ref{eqamp})), separately, and identified the obtained phases with 
$\delta_{\sigma}$ and $\delta_{BG}$, respectively.  
In actual calculation in addition to ${\cal K}_\lambda$ 
we also included the $t,u$-channel amplitudes 
$A(t,s,u)$, $A(u,t,s)$ and the $\rho$-meson 
contribution\footnote{$\rho$-meson contribution 
is included by using the Schwinger-Weinberg Lagrangian.}
in the ${\cal K}_{BG}$, and unitarized it 
following the $N/D$ method.
A subtraction constant necessary to make the $D$ convergent was fixed so as to give
$\delta_\sigma^{\rm th}=\delta_\sigma^{\rm phys}$ at $s=m_\sigma^2$. 
Our predicted $\delta_{BG}$ are seen to be almost
consistent\cite{rf11} with the phenomenological $\delta_{BG}$ in $I$=0 and $I$=2 channels
obtained in our phase shift analysis.\\
\underline{\it Relation to the other phase shift analyses}\ \ \ 
In our mehtod the ${\cal K}_\sigma$ and 
${\cal K}_{BG}$ (mainly coming from ${\cal K}_\lambda$)
are unitarized separately, and the total $S$-matrix is given by 
\begin{eqnarray}
S &\equiv & \frac{1+i\rho_1 {\cal K}}{1-i\rho_1 {\cal K}}=S_\sigma S_{\lambda};\ \ 
S_\alpha =\frac{1+i\rho_1 {\cal K}_\alpha}{1-i\rho_1 {\cal K}_\alpha}  ,
\ \ \ \ \  \alpha =\sigma ,\ \lambda ,
\label{eqS}
\end{eqnarray}
where the simple ${\cal K}$-matrix unitarization method is applied to  
${\cal K}_\lambda$, as well as ${\cal K}_\sigma$. 
The total ${\cal K}$-matrix defined above is given by
\begin{eqnarray}
{\cal K} &=& \frac{{\cal K}_\sigma +{\cal K}_{\lambda}}
{1-\rho_1^2{\cal K}_\sigma{\cal K}_{\lambda}} 
=\frac{(s-m_\pi^2)\frac{3(m_\sigma^2-m_\pi^2)}{f_\pi^2}}
{m_\sigma^2+(1-\frac{4m_\pi^2}{s})\frac{9(m_\sigma^2-m_\pi^2)^3}{(32\pi f_\pi^2)^2}-s}
\approx \frac{(s-m_\pi^2)\frac{3m_\sigma^2}{f_\pi^2}}{m^2-s},\ \ \ 
\label{eqIA}
\end{eqnarray}
where Adler zero factor $(s-m_\pi^2)$ appears. 
The resulting $S$-matrix has a pole at the light mass
$\approx m_\sigma$.  
The ${\cal K}$-matrix pole position is 
$s=m^2 \approx m_\sigma^2+(\frac{3m_\sigma^3}{32\pi f_\pi^2})^2$,
where the phase passes through 90 degrees.
In case of $m_\sigma \approx 600$ MeV, $m$ becomes $\approx 900$ MeV.
In this case, the $S$-matrix has a complex pole with mass $m_\sigma \approx 600$MeV, 
but the phase passes 90$^\circ$ at $m\approx 900$MeV.
This situation
of our phase shift analysis 
is common to the many other analyses motivated by L$\sigma$M. 
When the real Born amplitude, having 
one pole with Adler zero factor as in Eq.~(\ref{eqIA}),
\begin{eqnarray}
{\cal K} &=& (s-m_\pi^2)g^2/(m^2-s),
\label{eqK}
\end{eqnarray} 
is unitarized by ${\cal K}$-matrix or other methods, 
the light $\sigma$-meson pole in $S$-matrix is obtained generally:

In the analysis by Achasov et al.\cite{rf12}
they obtain large $m\approx 1000$MeV
through the fit to the experimental data, but 
their physical mass is very small as $m_\sigma =417 $ MeV.

In the analysis by Tornqvist et al,\cite{rf13} their ``Breit-Wigner" mass $m_{BW}$,
which corresponds to $m$ in our interpretation, is about  900 MeV, 
but the real part of the physical pole is $m_\sigma \approx 500$ MeV.   

In the analysis by Igi and Hikasa,\cite{rf14} the $N/D$-unitarization method is used. 
They use the large bare mass $m=m_\rho =770$ MeV.\footnote{
They include the crossed channel $\rho$- as well as $\sigma$- meson exchange.
They compare the phase shifts with and without $\sigma$-effect, 
and obtain the conclusion of $\sigma$-existence.
It should be noted that their $m_\sigma$ corresponds to our $m$. 
This mass is smaller than the other analyses motivated by L$\sigma$M, and thus 
the predicted phase shift passes through 90 degrees with somewhat smaller energy than 
the others. The real part of the $S$-matrix pole position, denoted as $m_\sigma$ 
in our notation, is not given in their paper.}

In the analysis by Au, Morgan and Pennington,\cite{rf15} the two-channel 
${\cal K}$-matrix of the similar structure as Eq.(\ref{eqK})\footnote{
They also try three pole fit, but 
the other two poles are far from the relevant energy region,
and this analysis is essentially equivalent to the above one-pole fit.  
} 
with a polynomial backgroud ${\cal K}_{\rm pol}$ was used. 
They obtained $m\approx 900$MeV, 
however, they only searched for the case of small $g$-value,  
which corresponds to narrow resonance like $f_0(980)$, and overlooked the 
possibility of wide resonance with large $g$.  
However,  
$\sigma$ is actually required to have wide width from Eq.(\ref{eqrel}).
Accordingly  
they were forced to introduce the large background contribution 
${\cal K}_{\rm pol}$ 
in order to fit experimental data\footnote{
Furthermore, in their fit 
the effect of virtual $K\bar K$-channel is extremely large in the energy region
far below the $K\bar K$-threshold. 
This fit seems to be unnatural, and necessarily to be corrected.
} 
Their result of no light 
$\sigma$-existence is simply to be due to this overlooking and is not correct.

\section{Concluding remark}
%

The observed values of mass and width 
of scalar $\sigma$-nonet 
($\sigma (600)$, $\kappa (900)$, $a_0(980)$ and $f_0(980)$) 
are consistent with the relation predicted by $SU(2)$ and $SU(3)$ L$\sigma$M. 
This fact suggests the linear representation of chiral
symmetry is realized in nature.
The strong cancellation between
$\delta_\sigma$ due to the $\sigma$ resonance 
and $\delta_{BG}$ introduced in our phase shift analysis
corresponds to the cancellation between $\sigma$-resonance amplitude and
repulsive $\lambda\phi^4$ amplitude in L$\sigma$M,
and this cancellation is guaranteed from the viewpoint of chiral symmetry.
The other phase shift analyses motivated by L$\sigma$M, 
leading to the $\sigma$-existence, 
also take into account this cancellation mechanism, explicitly or implicitly.
The reason of missing $\sigma$ in the phenomenological analysis 
by Au, Morgan and Pennington
is that they overlooked the possibility of wide resonance,
and did not pay attention to the cancellation mechanism of chiral symmtery breaking as depicted in Fig. 1, and thus 
their conclusion of no $\sigma$-existence is not correct.

Concerning on our analysis of $\delta^{I=0}$, Pennington made a 
criticism.\cite{rf16}
that the choice of $\delta_{BG}$ is completely arbitrary and
accordingly ``one can obtain more or less any set of Breit-Wigner parameters one likes 
for the $\sigma$."
However, The $\delta_{BG}$ has the clear physical  
origin in the compensating repulsive $\lambda\phi^4$ interaction, 
necessary from the viewpoint of chiral symmetry, and is not arbitrary.
Energy dependence of $\delta_{BG}$ predicted by using L$\sigma$M  
changes slightly depending on the unitarization methods, however,
the pole position of $\sigma$ is rather stable within the uncertainty given 
in our present estimate in Table I.
Thus, the criticism is not valid.

\end{document}